\documentclass[conference,10pt]{IEEEtran}
\IEEEoverridecommandlockouts

\usepackage[utf8]{inputenc}
\usepackage{color,graphicx,caption,subcaption}
\usepackage[cmex10]{amsmath}
\usepackage{amsfonts}
\usepackage[english]{babel}
\usepackage{amssymb,amsmath,mathtools}
\usepackage{enumerate}
\usepackage{algorithm}
\usepackage{algorithmic}
\usepackage{dblfloatfix}
\usepackage{bbm}
\usepackage{amsmath}
\usepackage{mathrsfs}
\usepackage{mathtools}
\usepackage{breqn}
\usepackage[english]{babel}
\usepackage{amsthm}
\usepackage{enumitem}
\usepackage{booktabs}
\usepackage{steinmetz}
\usepackage{cite}
\usepackage{color}
\usepackage{booktabs}
\usepackage{makecell}

\setlist[itemize]{leftmargin=*}
\setlist[enumerate]{leftmargin=*}

\newtheorem{Lemma}{Lemma}

\newcommand{\bff}[1]{\mathbf{#1}}

\newcommand{\lag}{\mathcal{L}}

\newcommand{\sqproof}{\hspace*{0em plus 1fill}\makebox{\hfill\ensuremath{\square}}}

\DeclareMathOperator*{\argmax}{argmax}

\title{\LARGE \bf
Joint User Scheduling and Power optimization in Full-Duplex Cells with Successive Interference Cancellation  
}
\author{Shahram Shahsavari, David Ramirez, and Elza Erkip 
\thanks{Authors are with the ECE Department of New York University. This work is supported in part by NSF Grant 1527750 and NYU WIRELESS. E-mails: \{shahram.shahsavari, david.ramirez, elza\}@nyu.edu.}}


\begin{document}
\bstctlcite{IEEEexample:BSTcontrol}
\maketitle
\thispagestyle{empty}
\pagestyle{empty}

\begin{abstract}
This paper considers a cellular system with a full-duplex base station and half-duplex users. The base station can activate one user in uplink or downlink (half-duplex mode), or two different users one in each direction simultaneously (full-duplex mode). Simultaneous transmissions in uplink and downlink causes self-interference at the base station and uplink-to-downlink interference at the downlink user. Although uplink-to-downlink interference is typically treated as noise, it is shown that successive interference decoding and cancellation (SIC mode) can lead to significant improvement in network utility, especially when user distribution is concentrated around a few hotspots. The proposed temporal fair user scheduling algorithm and corresponding power optimization utilizes full-duplex and SIC modes as well as half-duplex transmissions based on their impact on network utility. Simulation results reveal that the proposed strategy can achieve up to $95\%$ average cell throughput improvement in typical indoor scenarios with respect to a conventional network in which the base station is half-duplex.	
\end{abstract}

\section{Introduction}
Next-generation wireless networks require higher spectral efficiency to accommodate the increasing demand for data traffic. A full-duplex base station (BS) can simultaneously transmit in downlink (DL) and receive in uplink (UL), thus doubling the frequency reuse of the network which can lead to noticeable gains in spectral efficiency compared to conventional half-duplex (HD) networks.

Using the same frequency band for UL and DL transmissions introduces additional interference to the network. The power leakage between transmitter and receiver of the BS creates interference on UL reception known as self-interference. Although interference cancellation techniques are proposed to mitigate this interference at the BS \cite{duarte2012experiment}, it cannot be canceled completely due to impairments such as quantization and phase noise \cite{masmoudi2016self}. Additionally, the UL transmission creates interference on DL reception known as UL-to-DL interference. For succinctness, we refer to UL-to-DL interference as UDI.
To fully exploit capabilities of FD networks, a careful control of self-interference and UDI is required via signal processing techniques and careful network operation.

Analog and digital interference suppression techniques have been shown to achieve high levels of self-interference cancellation at the BS \cite{duarte2012experiment}. To suppress the impact of UDI, one approach, known as successive interference cancellation (SIC), is to decode the UDI signal first by treating the desired DL signal as noise, and then removing UDI to leave a clean channel for the desired DL signal \cite{el2011network}. SIC is most efficient when the UDI channel is strong since it is easier for DL user to decode and cancel a stronger UDI. Practical considerations of SIC are discussed in \cite{xu2008practical}. Altogether, these techniques motivate full-duplex as a feasible option for future networks \cite{goyal2015full}. 

Appropriate user scheduling and power optimization can mitigate interference in FD networks \cite{goyal2015full,goyal2017user}. To decrease the impact of the interference further, hybrid strategies can be used in which FD transmissions are scheduled whenever they can increase the utility and system falls back to HD otherwise. While a hybrid strategy is proposed in \cite{goyal2017user}, it does not consider the possibility of SIC at user devices.  

Fair partition of network resources combats user starvation and can improve quality of experience for users. Two popular forms of fairness considered in the literature are proportional fairness \cite{goyal2017user}, and temporal fairness \cite{shahsavari2015two}. Single-cell and multi-cell proportional fair schedulers are proposed for FD networks in \cite{goyal2014improving} and \cite{goyal2015full}, respectively, but without the possibility of SIC.

In this paper, we consider opportunistic temporal fair user scheduling and power optimization in a single-cell FD network. The proposed algorithm utilizes FD transmissions with or without SIC in an opportunistic manner to maximize weighted average of UL and DL rates. We demonstrate the effectiveness of SIC in FD networks especially when user distribution is concentrated around a few hotspots. Furthermore, we address the UL-DL traffic asymmetry common in wireless networks. Extensive simulation results demonstrate the performance of the proposed method.

\section{System Model}
We consider a single-cell time-slotted system consisting of a BS serving $K$ users over a single frequency channel. We use $U_i$ to denote user $i \in \{1,\ldots,K\}$. The BS is FD, hence it can activate one user in UL and one user in DL. Users are half-duplex (HD) either in UL or DL.

Suppose the BS activates two users one in UL and one in DL. There are two types of communication channels in this scenario: \textit{i}) the channels between users and the BS and \textit{ii}) the channel between the users, i.e., UDI channel. Let $g_i$ denote the complex channel coefficient between the BS and $U_i$. Similarly, let $h_{ij}$ denote the complex channel coefficient between $U_i$ and $U_j$. We consider a combination of large-scale (i.e. distance based path-loss and shadowing) and small-scale Rayleigh fading to model channel coefficients $g_i$ and $h_{ij}$. We let $G_i=|g_i|^2, \forall i$ and $H_{ij}=|h_{ij}|^2, \forall i, j$ denote the power gain corresponding to the channel coefficients $g_i$ and $h_{ij}$, respectively. We assume that the small scale channel coefficients are fixed during each time-slot and are independent across time-slots. Moreover, we assume that the scheduler has knowledge of $g_i$ and $h_{ij}~\forall~i,j \in \{1,\ldots,K\}$ at each time-slot. Estimates of $g_i$ can be obtained from reference signals in 3GPP LTE \cite{dahlman20134g}. To obtain $h_{ij}$, as in \cite{goyal2017user}, users can estimate using reference signals and feed the estimate back to the BS.

\begin{figure}[t]
 \centering \includegraphics[width=0.85\linewidth]{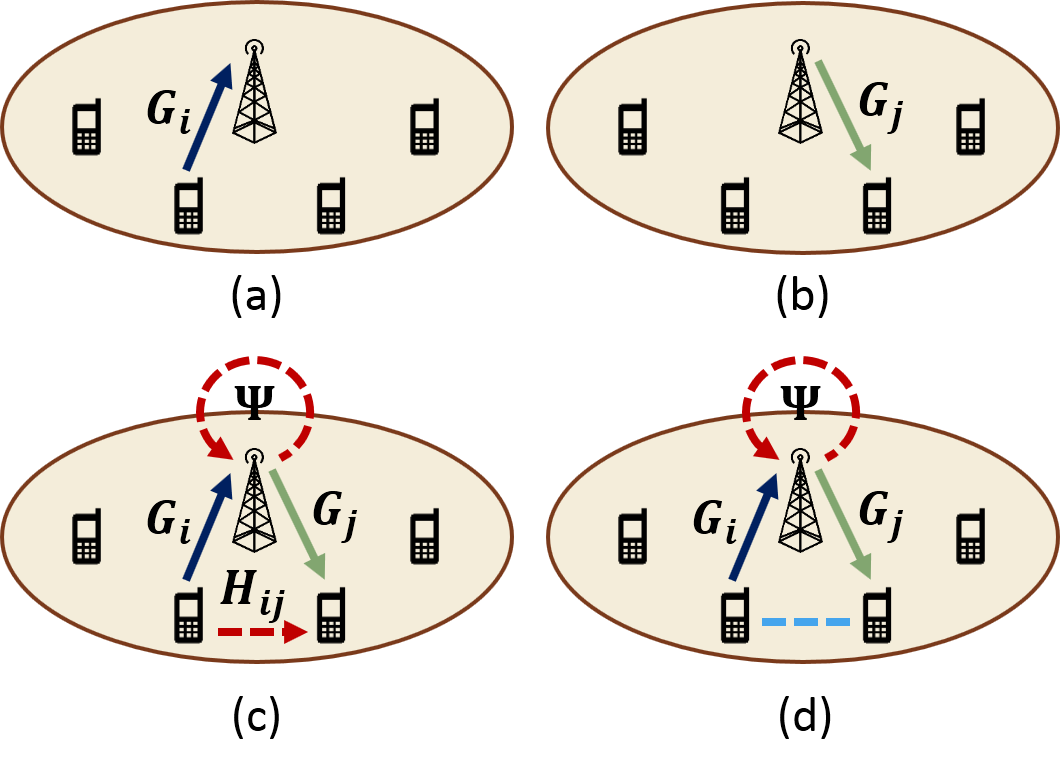}
 \caption{Graphical representation of communication modes. (a) HD mode in uplink, (b) HD mode in downlink, (c) FD mode, and (d) SIC mode }
 \label{fig:system-model}
\end{figure}

\subsection{Communication Modes}
\label{subsec:mode}
The network operates in one of the three communication modes described below. The communication mode defines the signal to interference and noise ratio (SINR), and the data rate is a function of the SINR detailed in Section \ref{sec:power-opt}.

\subsubsection{Half-duplex (HD) Mode}
In HD mode, the BS serves one user either in UL or DL; see Fig. \ref{fig:system-model} (a) and (b). Let $P_u$, $P_d$, $N_u$, and $N_d$ denote the transmit powers in UL and DL and noise powers in UL and DL, respectively. If $U_i$ is chosen for UL or DL in this mode then the SINRs are
\begin{align}
SINR_{UL}^{HD}(i)&=\frac{P_uG_i}{N_u},\label{eq:ul-hd}\\
SINR_{DL}^{HD}(i)&=\frac{P_dG_i}{N_d}.\label{eq:dl-hd} 
\end{align}

\subsubsection{Full-duplex (FD) Mode}
In FD mode, the BS serves one user in UL and another user in DL, see Fig. \ref{fig:system-model} (c). As a result, two interference signals are introduced. First, self-interference which is caused by the DL transmission interfering with the reception of the UL transmission. We denote $\Psi$ as the effective channel between output and input at the BS which is inversely proportional to the amount of self-interference cancellation at the BS.
Second interference is UDI caused by the UL transmission interfering at the DL user. 
We assume that residual self-interference and UDI are treated as noise in FD mode. Suppose that $U_i$ and $U_j$, $i\not= j$ are chosen for UL and DL, respectively, in FD mode, then the SINRs are 
\begin{align}
SINR_{UL}^{FD}(i,j)&=\frac{P_uG_i}{P_d\Psi+N_u},\label{eq:ul-fd}\\
SINR_{DL}^{FD}(i,j)&=\frac{P_dG_j}{P_uH_{ij}+N_d}.\label{eq:dl-fd}
\end{align}
Since FD mode treats UDI as noise, FD would be more efficient when $H_{ij}$ is small.

\subsubsection{Full-duplex with successive interference cancellation (SIC) Mode}
In SIC mode, the BS behaves as in FD mode while the UL rate is adjusted so that the DL user can utilize SIC to mitigate UDI; see Fig. \ref{fig:system-model} (d).
Suppose that $U_i$ and $U_j$, $i\not= j$ are respectively chosen for UL and DL in SIC mode, then the SINRs are 

\begin{align}
SINR_{UL}^{SIC}(i,j)&=\min\bigg\{\frac{P_uG_i}{P_d\Psi+N_u},\frac{P_u H_{ij}}{P_dG_j+N_d}\bigg\}, \label{eq:ul-sic}\\
SINR_{DL}^{SIC}(i,j)&=\frac{P_dG_j}{N_d}.\label{eq:dl-sic}
\end{align}

We note that performing SIC requires the DL user to decode the UL signal and cancel it from the received signal. On the other hand, the BS also needs to decode the UL signal. Therefore, UL user should ensure that the UL signal is decodable at both DL user and the BS. Equivalently, we can define the SINR of the UL as the minimum between the SINRs of UL-to-DL (UDI) and UL-to-BS channels as shown by \eqref{eq:ul-sic}. We note that the DL is interference free in this mode due to SIC, and $SINR_{UL}^{SIC}$ is not limited by the UDI channel if $H_{ij}$ is sufficiently large. 
In this case, SIC helps to cancel UDI without sacrificing the UL SINR. 

\subsection{Temporal fairness}
\label{subsec:fairness}
Without loss of generality, we assume that each user has UL and DL traffic to send and receive. Thus, we split each user $U_i$ into an UL user $U^{UL}_i$ and a DL user $U^{DL}_i$. We also define weights $w^{UL}_i$ and $w^{UL}_i$ for $U^{UL}_i$ and $U^{DL}_i$, respectively, such that $\sum_i (w^{UL}_i+w^{DL}_i)=1$. Let $a^{UL}_i$ and $a^{DL}_i$ be the the fraction of time-slots that $U^{UL}_i$ and $U^{DL}_i$ are activated in the long run (or air-time share of those users), respectively. In conventional HD networks where only one of the users $U^{UL}_i$, $U^{DL}_i, \forall i$ is chosen for each time-slot, the system is called weighted temporal fair if and only if $\forall i: a^{UL}_i=w^{UL}_i, a^{DL}_i=w^{DL}_i$ \cite{ShahsavariAK16}. When full-duplex transmissions are allowed, we call the system weighted temporal fair, if and only if $\forall i: a^{UL}_i\geq w^{UL}_i, a^{DL}_i \geq w^{DL}_i$. We note that while the summation of airtime shares is always one in HD networks, it can be up to two when full-duplex transmissions are allowed because of the possibility of activating up to two users per time-slot.

\section{Power optimization}
\label{sec:power-opt}
Power optimization can be used to control UDI and self-interference in FD and SIC modes. We note that the power optimization is trivial for HD mode in a single-cell scenario where each terminal transmits at full power. In this section, we solve the power optimization problem for a given pair of scheduled users for both FD and SIC modes.
In the next section, we consider joint temporal fair user scheduling, mode selection, and power optimization using the optimized power levels of this section. 

The goal of power optimization is to maximize the network utility for a given instantaneous power budget. A typical wireless network has more traffic demand in DL than UL. Considering the traffic demand asymmetry, we define the network utility function for each mode $X\in\{HD,FD,SIC\}$ as a weighted average of UL and DL rates in that mode, i.e., $R^X_N \triangleq \rho R^X_{DL}+(1-\rho)R^X_{UL}$. We note that in HD mode one of the UL or DL rates is zero. Moreover, $0 \leq \rho\leq 1$ and increasing $\rho$ increases the relative importance of DL to UL. We remark that $R^X_N$ is a function of: \textit{i}) scheduled user(s), \textit{ii}) communication mode, and \textit{iii}) DL and UL transmit powers. Suppose that users $i$ and $j$ are scheduled in UL and DL, respectively. We formulate power optimization problem as 

\begin{align}
&\boldsymbol{\Pi}^X:~(\tilde{P}^X_u,\tilde{P}^X_d)=\argmax_{P_u,P_d}~R^X_N (P_u,P_d) \notag\\
&\text{subject to:}~~0\leq P_u \leq P^{max}_u,~~0\leq P_d \leq P^{max}_d,\notag
\end{align}
where $P^{max}_u$ and $ P^{max}_d$ denote the maximum UL and DL transmit powers, respectively. Before solving problem $\bff{\Pi}^X$, we detail the relationship between rate and SINR detailed for different modes in Section \ref{subsec:mode}. We consider two rate models: \textit{i}) Shannon rate in which $R=\log_2(1+SINR)$ and \textit{ii}) LTE rate model which follows from practical discrete modulation and coding schemes, and for which $R$ is a staircase function of $SINR$. Next, we study $\bff{\Pi}^X$ for different rate models and for $X\in\{FD,SIC\}$.

\subsection{Shannon Rate}
In the following two subsections we consider $\bff{\Pi}^X$ assuming that $R^X_Y=\log_2(1+SINR^X_Y)$, where $X \in \{FD,SIC\}$ and $Y\in \{UL,DL\}$ and $SINR^X_Y$ is introduced in \eqref{eq:ul-fd}-\eqref{eq:dl-sic}.

\subsubsection{FD mode}
we can use the method of Lagrangian multipliers \cite[chapter~3]{bertsekas1999nonlinear} to solve $\bff{\Pi}^{FD}$.
 

\begin{Lemma}
\label{lem:optimal power-fd}
The optimal solution of problem $\bff{\Pi}^{FD}$ is one of the following candidates
\begin{enumerate}
\item $P^{[1]}_u=0$, $P^{[1]}_d=P^{max}_d$
\item $P^{[2]}_u=P^{max}_u$,$P^{[2]}_d=0$
\item $P^{[3]}_u=P^{max}_u$, $P^{[3]}_d=P^{max}_d$
\item The solutions of the system of quadratic equations \eqref{eq:fd-power-candid}
 
\end{enumerate}
\end{Lemma}

The proof is provided in Appendix \ref{app:lemma1}. To find the optimal solution of $\bff{\Pi}^{FD}$, it is sufficient to evaluate $R^{FD}_N$ at the candidate solutions in Lemma \ref{lem:optimal power-fd} and pick the one with maximum value. Finding every solution of the system of non-linear equations provided in \eqref{eq:fd-power-candid} can be computationally complex. However, extensive numerical examples in Section \ref{sec:simulations} show that even if we only consider the first three candidates in Lemma \ref{lem:optimal power-fd} (equivalently, using a two-level transmit power strategy), an optimal solution is often found.

\begin{figure*}[!t]

\begin{align}
&\begin{cases}
\rho G_j (P_d \Psi+P_u G_i+N_u)(P_d \Psi+N_u)-(1-\rho)P_uG_i \Psi (P_d G_j +P_u G_i+N_d)=0\\
\rho P_d G_j H_{ij} (P_d \Psi+P_u G_i+ N_u)-(1-\rho) G_i (P_dG_j+P_uH_{ij}+N_d)(P_u H_{ij}+N_d)=0
\end{cases} \label{eq:fd-power-candid}\\
&~~~\rho G_j(P_d G_j+P_u^{max}H_{ij}+N_d)(P_d G_j+N_d)-(1-\rho)P_u^{max}G_jH_{ij}(P_dG_j+N_d)=0 \label{eq:sic-power-candid1}\\
&~~~\rho G_j(P_d\Psi+P_u^{max}G_i+N_u)(P_d\Psi +N_u)-(1-\rho) P_u^{max} G_i\Psi (P_dG_j+N_d)=0 \label{eq:sic-power-candid2}
\end{align}
\hrule
\end{figure*}

\subsubsection{SIC Mode}
According to \eqref{eq:ul-sic} and \eqref{eq:dl-sic}, increasing UL transmit power does not decrease DL SINR due to UDI cancellation at DL user (as opposed to FD mode) while it increases the UL SINR. Thus, $\tilde{P}^{SIC}_u=P^{max}_u$ is optimal, and problem $\bff{\Pi}^{SIC}$ reduces to finding the optimal DL transmit power. 
As in FD mode, we use the method of Lagrangian multipliers to find the optimum DL transmit power.


\begin{Lemma}
\label{lem:optimal power-sic}
The optimal DL power in problem $\bff{\Pi}^{SIC}$ is one of the following candidates
\begin{enumerate}
\item $P^{[1]}_d=0$
\item $P^{[2]}_d=P^{max}_d$
\item $P^{[3]}_d=(H_{ij}N_u-G_iN_d)/(G_iG_j-\Psi H_{ij})$
\item $P^{[4]}_d$ and $P^{[5]}_d$, the solutions of the quadratic equation \eqref{eq:sic-power-candid1}
\item $P^{[6]}_d$ and $P^{[7]}_d$, the solutions of the quadratic equation \eqref{eq:sic-power-candid2} 
\end{enumerate}
\end{Lemma}

The sketch of proof is provided in Appendix \ref{app:lemma2}. Note that Lemma \ref{lem:optimal power-sic} provides every candidate for the optimal DL transmit power in $\bff{\Pi}^{SIC}$. Therefore, it is sufficient to evaluate objective function $R^{SIC}_N$ at these candidates (with $\tilde{P}^{SIC}_u=P^{max}_u$) and pick the one with the maximum value. Unlike the FD mode candidates, evaluating the SIC mode candidates are not computationally taxing. 

\subsection{LTE rate model}
\label{subsec:lte-power-opt}
The discrete modulation and coding schemes used in practical LTE systems, result in rate being a staircase function of SINR. Since this function is not continuous, solving $\bff{\Pi}^X$ becomes harder.  
We consider multiple discrete power levels for UL and DL and perform a naive search and select the power levels with the largest network utility. As in the Shannon rate case, it is sufficient to use $\tilde{P}^{SIC}_u=P^{max}_u$ and search only for $P_d$ for the SIC mode. Numerical results in Section \ref{sec:simulations} show that even a two-level power search strategy (on-off strategy) can lead to considerable gains in the network utility without adding a computational burden.

\section{Joint User Scheduling, Mode Selection and Power Optimization (JSMP)}
\label{sec:jsmp}
In the previous section, we discussed power optimization to maximize network utility for different modes assuming that a pair of users are scheduled. In this section, we address the scheduling problem, namely which user(s) should be scheduled at each time-slot so as to maximize the long-term average network utility while guaranteeing weighted temporal fairness. Because of the fairness constraint, the optimal strategy is not necessarily to activate the user(s) with the best utility at each time-slot. Furthermore, presence of full-duplex transmissions complicates the problem further, because activating two users together will increase airtime share of both users. Therefore, it is difficult to obtain the optimal temporal fair scheduler and we propose a sub-optimal (heuristic) scheme called \textit{joint user scheduling, mode selection, and power optimization (JSMP)} that guarantees weighted temporal fairness.
We define virtual user $V_{ij}$ as the pair of users $U^{UL}_i$ and $U^{DL}_j$ where $i,j\in\{0,1,2,\ldots K\}$. When $i=0$ or $j=0$, there is no user in UL and DL, respectively. Moreover, we assume $i\not=j$ in $V_{ij}$ since users are HD. It is clear that $V_{ij}$ can represent any individual user either in UL or DL in HD mode, and any pair of users one in each direction for FD and SIC modes. 
In JSMP, the following steps are considered for every time-slot:  

\begin{itemize}[label={}]

\item{\textbf{Step 1}}: 
We select the best mode and optimized powers for each virtual user $V_{ij}$. If $i=0$, the mode can only be HD and the optimized power is $P_d=P_d^{max}$. Similarly, for $j=0$ the mode is HD and $P_u=P_u^{max}$. If $i$ and $j$ are both non-zero, then we find the best mode and transmit powers for $V_{ij}$ as follows.
\begin{align}
&(\tilde{P}^X_u,\tilde{P}^X_d)=\argmax_{P_u,P_d}~R^X_N (P_u,P_d,V_{ij}), ~X\in \mathcal{X}, \label{eq:jsmp-power-opt-vij} \\
& X_{ij}^*=\argmax_{X\in \mathcal{X}}~R^X_N (\tilde{P}^X_u,\tilde{P}^X_d,V_{ij}),  \label{eq:jsmp-best-mode-vij} \\
&\tilde{P}^{ij}_u=\tilde{P}^{X_{ij}^*}_u,\tilde{P}^{ij}_d=\tilde{P}^{X_{ij}^*}_d,  \label{eq:jsmp-best-power-vij}
\end{align}
where $X_{ij}^*$ is the best mode for $V_{ij}$ and $\tilde{P}^{ij}_u$ and $\tilde{P}^{ij}_d$ are the optimal transmit powers in UL and DL, respectively. $\mathcal{X}$ denotes the set of available modes which can be a subset of $\{HD,FD,SIC\}$ in this paper. We note the dependence of $R^X_N$ on $V_{ij}$ is stated explicitly in \eqref{eq:jsmp-power-opt-vij} as opposed to the power optimization problems formulated in Section \ref{sec:power-opt} where we assumed the scheduled users are given.

\item{\textbf{Step 2}}:
We assign to virtual user $V_{ij}$ the utility $q_{ij}$ which is the network utility obtained by activating $V_{ij}$ in mode $X_{ij}^*$ with transmit powers $\tilde{P}^{ij}_u$ and $\tilde{P}^{ij}_d$, i.e., $q_{ij}=R^{X_{ij}^*}_N(\tilde{P}^{ij}_u,\tilde{P}^{ij}_d,V_{ij})$.

\item{\textbf{Step 3}}:
We define utility $Q^{UL}_i=\max_j q_{ij}$ for UL user $U^{UL}_i$. We also define $j^*_i=\argmax_j q_{ij}$ as the index of the best partner for each UL user $U^{UL}_i$. Similarly, we define $Q^{DL}_j=\max_i q_{ij}$ and $i^*_j=\argmax_i q_{ij}$ as the utility and the index of the best partner for each DL user $U^{DL}_j$, respectively.

\item{\textbf{Step 4}}:
The utilities $Q^{UL}_i$ and $Q^{DL}_j$, $i,j=1,2,\ldots N$ are used by a weighted temporal fair HD scheduler whose goal is to maximize the long-term average network utility subject to fairness constraints. The cell-level HD scheduler proposed in \cite{shahsavari2015two} is adopted for this purpose. 
This scheduler picks one user among $U^{UL}_i$ and $U^{DL}_j$, $i,j=1,2,\ldots N$, opportunistically based on their utilities $Q^{UL}_i$ and $Q^{DL}_j$ for each time-slot while guaranteeing to pick user $U^{UL}_i$ and $U^{DL}_j$ in $w^{UL}_i$ and $w^{DL}_j$ fraction of the time-slots, respectively. Whichever user that the HD scheduler chooses in this step, that user and its best partner are scheduled using the optimal transmit powers and the best mode obtained in previous steps. 
\end{itemize}
We remark that JSMP guarantees weighted temporal fairness, i.e., $\forall i: a^{UL}_i\geq w^{UL}_i, a^{DL}_i \geq w^{DL}_i$ since each user $U^{UL}_i$ ($U^{DL}_i$) is either being picked directly by the HD scheduler (step 4) in $w^{UL}_i$ ($w^{DL}_i$) fraction of the time-slots, or as the best partner of another user. We note that power optimization and mode selection are performed to increase instantaneous network utility, and the goal of user scheduling is to maximize the expected value (long-term average) of the same network utility function while guaranteeing weighted temporal fairness.

\section{Simulation Results}
\label{sec:simulations}
We consider a single square remote radio head (RRH) Hotzone cell \cite{3gpp-channel} with dimensions $40$m $\times$ $40$m having a BS in the center with $K=6$ active users distributed around the BS with an exclusion of central disk with radius $r_{min}=5$m. This model describes indoor scenarios such as a floor in an office building \cite{3gpp-channel}. Two different user distribution models are considered: \textit{i}) uniform user distribution and \textit{ii}) hotspot model with $N_h$ hotspots inside the cell. We assume that $K/N_h$ users are distributed uniformly within a circle of radius $10$m around each hotspot. $N_h=0$ represents the case where there is no hotspot and the users are distributed uniformly at random within the cell. We consider ordinary temporal fairness that is $\forall i: w^{UL}_i=w^{DL}_i=\frac{1}{2N}$. Table \ref{tab:sim-param} lists relevant simulation parameters adopted from indoor RRH/Hotzone scenario introduced in \cite{3gpp-channel}. LTE rate model is adopted from \cite[Table 7.2.3-1]{3gpp-rate}, where there are 15 combinations of modulation and coding schemes. Assuming non-line of sight (NLOS) scenario, $P^{max}_u$ and $P^{max}_d$ are designed such that the UL and DL average SNR of $5$ dB is achievable in HD mode on the boundary of the cell, i.e. at $d=40\sqrt{2}$m.

\begin{table}[h]
\centering
\begin{tabular}{ll} \toprule
$\bf Parameter$ 				 & $\bf Value$    \\ \midrule
Bandwidth      					 & $10$ MHz                      \\
Noise spectral density     	     & $-174$ dBm/Hz                  \\ 
Noise figure                     & BS:8 dB, user: 9 dB\\
Number of hotspots               & $0,1,2,3$                      \\ 
Self-interference cancellation   & $80$ dB, $100$ dB \\

\makecell[l]{Log-normal shadowing \\ standard seviation}  & LOS: 3 dB, NLOS: 4 dB \\
Pathloss in dB ($d$ in km)      	 & \makecell[l]{LOS: $89.5+16.9\log_{10}(d)$\\ NLOS: $147.4+43.3\log_{10}(d)$}   \\ 
Small-scale fading model			 & Rayleigh fading \\
Rate model                 & Shannon/LTE                 \\\bottomrule

\end{tabular}
\caption{Simulation parameters.}
\label{tab:sim-param}
\end{table}
\subsection{Two-level (binary) power strategy}
\label{subsec:binary-power-fd}
We first evaluate the performance of two-level (binary) power strategy for problem $\bff{\Pi}^{FD}$. For each network instance, users are uniformly distributed within the cell. We assume $\Psi^{-1}=100$ dB self-interference cancellation at the BS. At each instance and for every pair of users, we evaluate $R^{FD}_N$ with $\rho=0.5$ at UL and DL transmit powers belonging to $\{0,P_u^{max}\}\times \{0,P_d^{max}\}$ and pick the pair of powers which leads to the highest objective value. As a benchmark, we also perform an exhaustive search for the optimal powers in $\{0,0.01P_u^{max},0.02P_u^{max},\ldots,P_u^{max}\}\times 
\{0,0.01P_d^{max},0.02P_d^{max},\ldots,P_d^{max}\}$. Fig. 2 (a) and (b) depict the empirical CDF of the objective value corresponding to the binary power strategy and the benchmark exhaustive search. We observe that for Shannon rate, the binary strategy is almost as good as the exhaustive search. This implies that considering the first three candidates discussed in Lemma \ref{lem:optimal power-fd} leads to a nearly optimal solution of $\bff{\Pi}^{FD}$. While there is a gap between two curves in LTE rate case, we show that the binary strategy is sufficient to achieve acceptable gains in the average network throughput. 
\begin{figure}[h]
 \centering \includegraphics[width=\linewidth, height=3.5cm]{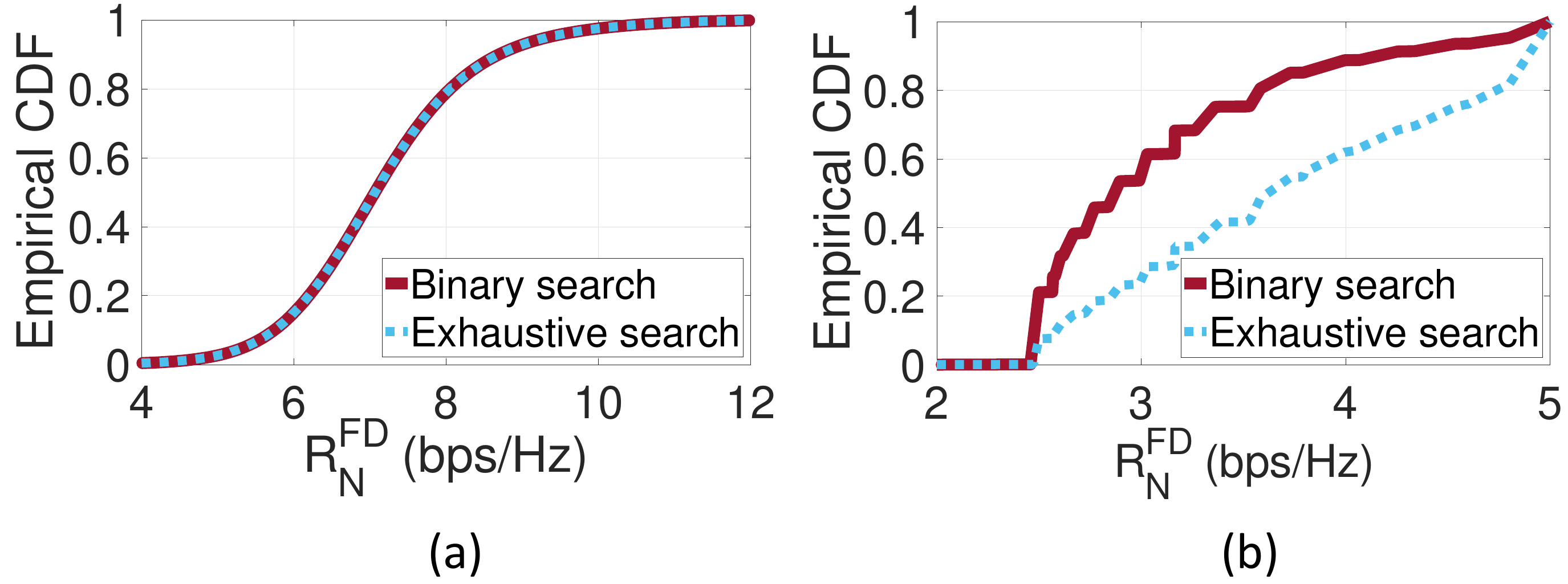}
 \caption{Performance of computationally simple binary power strategy against an exhaustive search for (a) Shannon rate and (b) LTE rate.}
 \label{fig:utility-gain}
\end{figure}
\subsection{Throughput gain}
\label{subsec:utility-gain}
In this section, we evaluate the performance of the proposed JSMP for a variety of parameters. First, we define three scenarios in terms of available underlying modes: \textit{i}) \textit{HD} scenario in which the BS and user devices are all HD. This is the benchmark scenario in which we run JSMP with $\mathcal{X}=\{HD\}$. \textit{ii}) \textit{HD+FD} scenario in which the BS is full-duplex but user devices cannot perform SIC. We run JSMP with $\mathcal{X}=\{HD,FD\}$ in this scenario. \textit{iii}) \textit{HD+FD+SIC} scenario in which the BS is FD and user devices can perform SIC. We run JSMP with $\mathcal{X}=\{HD,FD,SIC\}$ in this scenario. We consider $\rho=0.5$ in this example, i.e., network utility is assumed to be the average cell throughput. Moreover, we use binary power strategy discussed in Section \ref{subsec:binary-power-fd} for FD mode and the results of Lemma \ref{lem:optimal power-sic} for SIC mode when using Shannon rate in JSMP. For LTE rate we use binary power strategy for both FD and SIC modes as mentioned in Section \ref{subsec:lte-power-opt}. Fig. \ref{fig:utility-gain} illustrates the average cell throughput percentage gain of the scenarios \textit{HD+FD} and \textit{HD+FD+SIC} over the benchmark scenario \textit{HD}.

We consider $N_h\in \{0,1,2,3\}$ and $\Psi^{-1}\in \{80, 100\}$ dB and use both Shannon and LTE rate models. The gains over HD are significantly higher for LTE rate model because in a single-cell scenario there is no interference for the benchmark HD scenario and, more importantly, Shannon rate increases with $SINR$. Therefore, rates in the HD scenario are high which leads to less improvement. On the other hand, the LTE rate model has a finite maximum rate. The difference between $HD+FD+SIC$ and $HD+FD$ scenarios represents the gain of SIC in Fig. \ref{fig:utility-gain}. We observe that this gain is higher when $N_h=1$, because users are closer to each other in this case hence UDI is strong and SIC results in significantly higher rates. We also observe that the gain of scenario $HD+FD$ is small in this case, because without SIC strong UDI cannot be canceled. As the user distribution gets more uniform, i.e., for $N_h=2,3$, the marginal gain of SIC decreases and the performance of $HD+FD$ scenario gets better due to less severe UDI. Finally, note that both $HD+FD$ and $HD+FD+SIC$ scenarios lead to approximately the same gains when the users are uniformly distributed ($N_h=0$). Results reveal that higher self-interference cancellation leads to higher performance gains due to higher UL rates as expected.

\begin{figure}[h]
  \centering \includegraphics[width=0.6\linewidth]{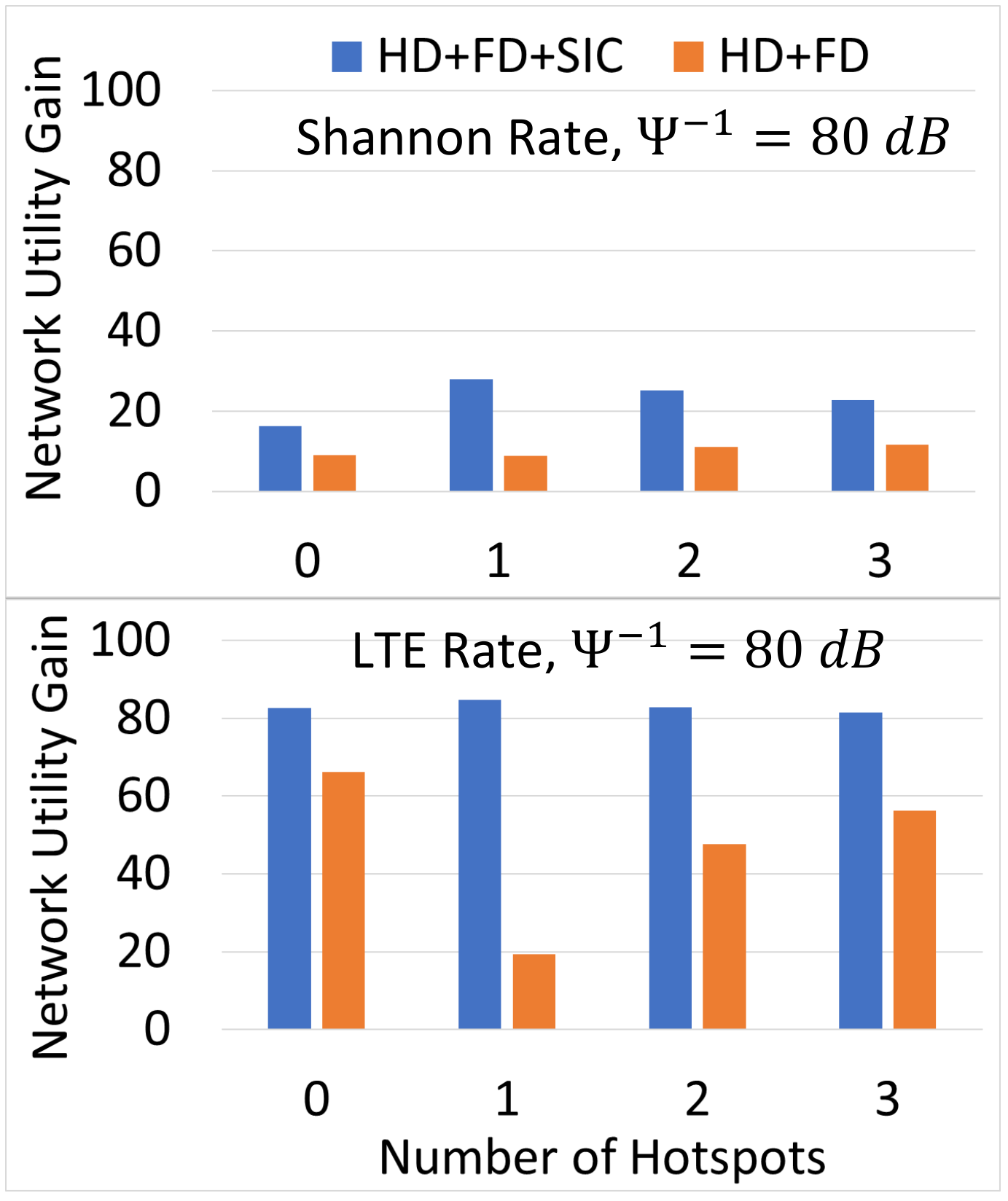}
  \centering \includegraphics[width=0.6\linewidth]{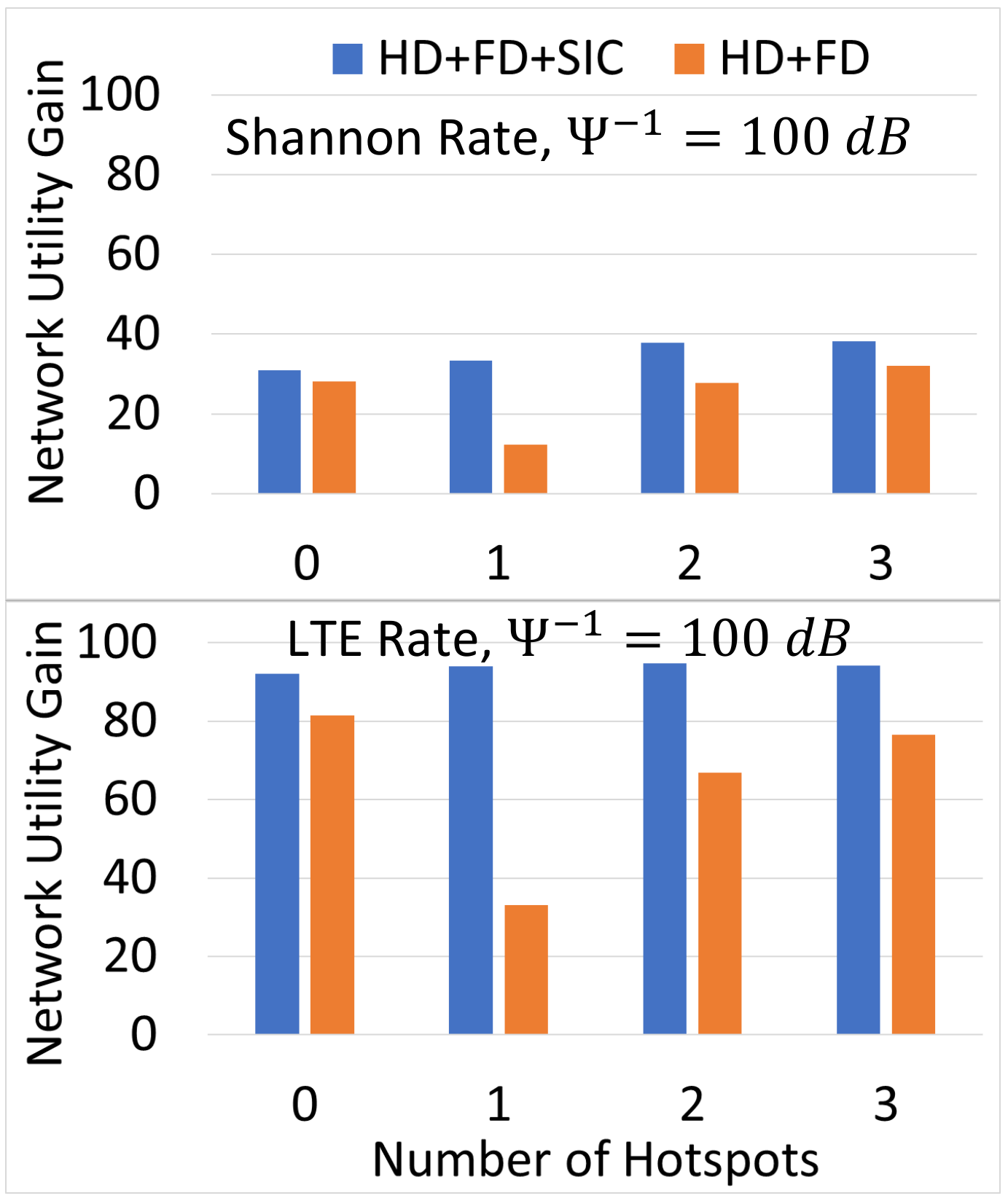}
 \caption{Network utility percentage gain of considering all three modes (first bar in each pair) or only HD and FD modes (second bar in each pair) relative to the HD mode.}
 \label{fig:utility-gain}
\end{figure}

\subsection{Traffic Asymmetry}
There is more traffic in DL compared to UL in typical wireless networks. One way to address traffic asymmetry is to increase parameter $\rho$ introduced in Section \ref{sec:power-opt}. We note that as $\rho$ gets higher, DL rate has higher impact on network utility which leads to higher priority being given to DL rate in mode selection, power optimization, and scheduling. 
We assume uniform user distribution, LTE rate model, $\Psi^{-1}=80$ dB, and run JSMP considering three scenarios introduced in Section \ref{subsec:utility-gain} for every $\rho\in \{0.3,0.5,0.7\}$.  To measure traffic imbalance, we define traffic asymmetry parameter $\gamma \triangleq \mathbb{E}(R_{DL})/\mathbb{E}(R_{UL})$. 
Table \ref{tab:asymmetry} details $\gamma$ as a function of $\rho$ for different scenarios. While increasing $\rho$ increases $\gamma$ in all the scenarios as expected, it has a greater effect on $\gamma$ in scenarios $HD+FD$ and $HD+FD+SIC$. This is because mode selection and power optimization are trivial in scenario $HD$, whereas in the other two scenarios, increasing $\rho$ allows for finer tuning of the relative DL and UL rates.
\begin{table}[h]
\centering
\scalebox{1}{
\begin{tabular}{lccc} \toprule
\textbf{Scenario}	& $\rho=0.3$  & $\rho=0.5$  & $\rho=0.7$ \\ \midrule
$HD$     		    & $0.97$      & $1$         & $1.02$	 \\
$HD+FD$     		& $0.66$      & $0.82$      & $1.51$	 \\
$HD+FD+SIC$         & $0.85$      & $1.07$      & $1.2$	     \\ \bottomrule
\end{tabular}}
\caption{Traffic asymmetry parameter $\gamma$ as a function of $\rho$}
\label{tab:asymmetry}
\end{table}
 
\section{Conclusions}
We have proposed a heuristic algorithm called JSMP for joint temporal fair user scheduling, mode selection, and power optimization in a single-cell full-duplex network. Simulating JSMP in a practical indoor scenario has revealed considerable gains in average cell throughput compared to conventional half-duplex networks. 
Furthermore, we have shown that while full-duplex transmissions do not lead to significant gains when users are located closely, successive interference cancellation can be performed to recoup full-duplex throughput gain. We have also addressed typical uplink-downlink traffic asymmetry in wireless networks by controlling the parameters of JSMP. Future work should aim to find the optimum joint design of scheduler, mode selection, and power optimization and extend the ideas to a multi-cell scenario.

\begin{appendices}
\section{Proof of Lemma 1}
\label{app:lemma1}
It is straightforward to show that the objective function of $\bff{\Pi}^{FD}$ is continuous and upper-bounded over the feasible set which is a closed and bounded set (i.e. compact). Therefore, per the extreme value theorem, there exists at least one optimal solution $(P_u, P_d)$ taking the global maximum
over the feasible set. It can be shown that $\bff{\Pi}^{FD}$ is equivalent to minimizing $(1+SINR^{FD}_{DL})^{-\rho} (1+SINR^{FD}_{UL})^{\rho-1}$ subject to the same constraints as in $\bff{\Pi}^{FD}$. The Lagrangian is
\begin{equation}
\begin{split}
\lag(P,\lambda,\mu)= (1+SINR^{FD}_{DL})^{-\rho} (1+SINR^{FD}_{UL})^{\rho-1}\\ +\lambda_u(P_u-P_u^{max})-\mu_uP_u+\lambda_d(P_d-P_d^{max})-\mu_dP_d, 
\end{split}
\notag
\end{equation}
where $P\triangleq [P_u,P_d]$ is the power vector and $\lambda\triangleq [\lambda_u,\lambda_d]$ and $\mu\triangleq [\mu_u,\mu_d]$ are the Lagrangian multipliers associated with the power constraints. Any optimal solution satisfies first order necessary conditions, i.e.
\begin{align}
&\nabla_P \lag(P,\lambda,\mu)=0,\label{eq:kkt-gradient-FD}\\
&0 \leq P_u \leq P^{max}_u, 0 \leq P_d \leq P^{max}_d \label{eq:kkt-primal-feasibility-FD}\\
&\lambda_u (P_u-P_u^{max})=0,~ \mu_u P_u=0, ~~~ \label{eq:kkt-complementary-slackness-FD-u}\\
&\lambda_d (P_d-P_d^{max})=0,~ \mu_d P_d=0, ~~~ \label{eq:kkt-complementary-slackness-FD-d}\\
&\lambda_u \geq 0, \lambda_d \geq 0, \mu_u \geq 0, \mu_d \geq 0. \label{eq:kkt-dual-feasibility-FD}
\end{align}
Solving \eqref{eq:kkt-gradient-FD}-\eqref{eq:kkt-dual-feasibility-FD} leads to every possible candidate for optimality. Since $\lambda_u, \lambda_d, \mu_u,$ and $\mu_d$ can each be either zero or positive, there are $16$ possible cases for a solution. Any $\lambda_u>0,\mu_u>0$ or $\lambda_d>0,\mu_d>0$ contradicts \eqref{eq:kkt-complementary-slackness-FD-u} and \eqref{eq:kkt-complementary-slackness-FD-d}, hence $7$ of $16$ possible cases are infeasible. The case $\mu_u>0,\mu_d>0,\lambda_u=\lambda_d=0$ leads to zero power for UL and DL which is obviously not optimal. The case $\mu_u>0,\lambda_d>0,\lambda_u=\mu_d=0$ leads to $P^{[1]}_u=0$, $P^{[1]}_d=P^{max}_d$. The case $\lambda_u>0,\mu_d>0,\mu_u=\lambda_d=0$ leads to $P^{[2]}_u=P^{max}_u$,$P^{[2]}_d=0$. The case $\lambda_u>0,\lambda_d>0,\mu_u=\mu_d=0$ leads to $P^{[3]}_u=P^{max}_u$, $P^{[3]}_d=P^{max}_d$. It can be shown that remaining 5 cases along with \eqref{eq:kkt-gradient-FD} lead to the system of quadratic equations provided in \eqref{eq:fd-power-candid}. \sqproof

\section{Proof of Lemma 2}
\label{app:lemma2}
The proof is similar to proof of Lemma \ref{lem:optimal power-fd} except there is only one optimization variable, $P_d$, and the UL SINR expression \eqref{eq:ul-sic} is more complicated than \eqref{eq:ul-fd} used in Lemma \ref{lem:optimal power-fd}. It can be shown that when $P_d>P^{[3]}_d$ the first term in \eqref{eq:ul-sic} is larger, and for $P_d<P^{[3]}_d$ the second term is larger. Thus, we consider three possible scenarios: \textit{i}) $0\leq P^{[3]}_d\leq P_d^{max}$, \textit{ii}) $P^{[3]}_d<0$, and \textit{iii}) $P^{[3]}_d>P_d^{max}$. Applying necessary conditions to the problem in these three scenarios leads to the candidates provided in Lemma \ref{lem:optimal power-sic}. \sqproof

\end{appendices}

\bibliographystyle{IEEEtran}
\bibliography{FD}

\end{document}